\documentclass[runningheads]{llncs}
\usepackage{graphicx}

\begin{document}
\title{Healthcare Data Governance, Privacy, and Security - A Conceptual Framework}
%
%
\author{Amen Faridoon\inst{1,2}\orcidID{0000-0002-2129-4218} \and
M. Tahar Kechadi\inst{1,2} }
\authorrunning{A. Faridoon, M. Kechadi}
%
\institute{University College Dublin, Dublin, Ireland \and
The Insight Centre for Data Analytics, Dublin, Ireland 
\email{amen.faridoon@ucdconnect.ie}\\
\email{tahar.kechadi@ucd.ie}}
\maketitle 
\begin{abstract}
The abundance of data has transformed the world in every aspect. It has become the core element in decision-making, problem-solving, and innovation in almost all areas of life, including business, science, healthcare, education, and many others. Despite all these advances, privacy and security remain critical concerns of the healthcare industry. It’s important to note that healthcare data can also be a liability if it is not managed correctly. This data mismanagement can have severe consequences for patients and healthcare organisations, including patient safety, legal liability, damage to reputation, financial loss, and operational inefficiency. Healthcare organisations must comply with a range of regulations to protect patient data. We perform a classification of data governance elements/components in a manner that thoroughly assesses the healthcare data chain from a privacy and security standpoint. After deeply analysing the existing literature, we propose a conceptual privacy and security-driven healthcare data governance framework.

\end{abstract}

\section{Introduction}
\label{sec:intro}
Digital data collection has undergone a transformative journey since its inception. It started with manual data entry and simple electronic databases in the mid $ 20^{th}$ century. Since then, the process has gained momentum with the proliferation of computers and the internet. The $21^{st}$ century has witnessed an explosion in data sources, from social media and IoT devices to advanced sensors and machine-generated data. This influx prompted the development of advanced data collection techniques, including big data analytics, machine learning, and AI-driven insights. Similarly, healthcare data collection spans several centuries, evolving from rudimentary record-keeping to sophisticated digital systems. The $ 20^{th}$ century saw the rise of electronic health records (EHRs), allowing for more efficient patient information storage and retrieval. With the advent of the $21^{st}$ century, technological advancements enabled the integration of various data sources, such as wearable devices and mobile apps, contributing to the growth of comprehensive and real-time healthcare data collection. This evolution has significantly enhanced healthcare delivery, research, and policy-making by providing insights into disease patterns, population health trends and treatment effectiveness.

However, this evolution has also raised concerns about privacy, security, and ethical considerations, leading to stringent data protection regulations to balance innovation with individual rights. Healthcare data often contains sensitive and personal patient information, including medical history, diagnoses, treatments, prescriptions, test results, social security numbers, and home addresses, making it highly valuable and vulnerable to misuse and theft. Table \ref{table:EHR} represents the types of information stored in the electronic health records and their sensitivity level. Data privacy and security are essential for protecting patient rights, preventing identity theft and cyber-attacks, maintaining trust in the healthcare system, ensuring continuity of care, and complying with legal and ethical standards. Unfortunately, healthcare organizations remain a popular target for hackers due to the sensitive nature of the data they possess. For instance, in 2020, there were 616 reported data breaches affecting healthcare organizations in the United States, according to the HIPAA Journal. Recently, in January 2023, LifeBridge Health in Maryland exposed the personal and medical information of approximately 1.4 million patients, including names, birth dates, medical diagnoses, and treatment information. These data breach cases demonstrate the insufficient implementation and management of security and privacy measures.
\begin{table*}
 \caption{Types of Information Present in Electronic Health Records and Their Sensitivity Level}
 \centering
 \resizebox{\textwidth}{!}{
 \begin{tabular*}{32pc}{@{\extracolsep{\fill}}| p{3.0cm}| p{6.8cm}| p{3.2cm}| @{}}
 \hline
Common Features & Description & Sensitivity Level \\ 
\hline
 Entity Identifiers & Personal identification information such as; name, address, email, phone number etc. & Identifiable \\ 
 \hline
 Demographic Information & Classification of a person in a specific group such as; age, education, gender, area etc. & Quasi-Identifiable \\ 
 \hline
 Clinical Records & Patient's medical history include treatments, diagnoses, medication,n etc. & Quasi-Identifiable and sensitive \\ 
 \hline
 Medical Biometrics & Patient's physical health-related information such as; blood pressure, heart rate, X-Ray, test reports etc. & sensitive \\ 
 \hline
 Mental Health Information & Patient's psychological related information such as; sleep problems, dietary information, psychosocial issues etc. & sensitive \\ 
 \hline
 Activities and lifestyle & Person's physical activities and their lifestyle-related information such as; physical activities, nutrition, exercise plans, etc. & sensitive \\ 
 \hline
 Financial Information & Person's financial data such as; health insurance, billing, reimbursements, financial class etc. & Quasi-Identifiable and sensitive \\ 
 \hline
 IoT and Wearable Data & Person's wearable and monitoring data include healthcare wearable devices, healthcare IoT devices, sensors data, etc. & Quasi-Identifiable and sensitive \\ 
 \hline
\end{tabular*}
 }
\label{table:EHR}
\end{table*}

Inadequate management of data can lead to potential legal responsibilities. Healthcare organizations
need to proactively safeguard patient information through solid security and privacy protocols to
counter various risks. Considering the sensitivity, usability, and multiple access to electronic
health records, healthcare institutions should adopt preemptive data governance frameworks. These
frameworks ensure operational efficiency without compromising the confidentiality of data, even from
internal individuals with authorised system access who might have malicious intent. The concept of
Electronic Health Record (EHR) Governance was first introduced by the Institute of Medicine (IOM) in
their 2003 report titled "Key Capabilities of an Electronic Health Record System"
\cite{aspden2004key}. This report emphasized the necessity of a systematic approach to managing EHR
systems, including the requirement of governance structures that could supervise the development,
implementation and maintenance of these systems. Similarly, the report "Nationwide Privacy and
Security Framework for Electronic Exchange of Individually Identifiable Health Information" was
issued by the US Department of Health and Human Services (HHS) in 2008
\footnote{https://digital.ahrq.gov/health-it-tools-and-resources/health-it-bibliography/privacy-and-security/nationwide-privacy-and}. This
report accentuates the need for governance frameworks to guarantee the secure and efficient
utilisation of Electronic Health Records. Organisations like the Office of the National
Coordinator for Health Information Technology (ONC) have created guidelines and frameworks to
assist EHR governance. These resources concentrate on aspects such as data privacy, data security,
data quality, and data interoperability. Data governance encompasses the comprehensive
administration of data within an organization. It entails assigning roles and duties for data
management, establishing regulations and protocols for ensuring data accuracy, confidentiality,
protection, and adherence to regulations, and setting up mechanisms to uphold these regulations. The
primary goal of data governance is to establish well-defined principles/guidelines for gathering,
retaining, accessing, and utilizing data to optimize its usefulness while minimising potential
hazards like data breaches and legal violations.

\paragraph{Contributions:}
We have managed to conduct a wide review of the concerns related to data privacy and security by
focusing on or analysing data governance activities in the healthcare industry. After deeply
analyzing the literature, we categorised data governance activities presented in section
\ref{sec:DGCF}. The primary goals of this critical evaluation are as follows: 
\begin{itemize}
 \item Study the existing state-of-the-art healthcare data governance models/systems.
 \item Categorize the data governance activities/elements in such a way that deeply analyses the
   healthcare data chain with the privacy and security perspective. 
 \item Propose a privacy and security-driven conceptual healthcare data governance framework. 
\end{itemize}

\section{Existing Healthcare Data Governance Models}
\label{sec:SHDGM}

Delving into Helen Nissenbaum’s approach (explains privacy as more than just a right to secrecy or
control; instead, it is about the suitable flow of private information within specific social
contexts) to privacy (2010), the study \cite{winter2019big} presenting a fascinating work for
accessing data governance in a specific context of private health data. This study examines
the use scenario involving the Royal Free Trust and Alphabet’s AI Venture DeepMind Health initiative. It
sheds light on the clashes among the partners concerning crucial aspects, particularly the
governance systems, objectives and gain attained via initiative. Researchers emphasize the
intricacies of governing PHI data to foster healthcare innovation and advancement while safeguarding
privacy and serving the public benefit. In a study \cite{winter2017investigating}, the authors
discovered six interconnected yet distinct models for governing Personal Health Information (PHI) by
focusing on what types of value are possibly afforded by PHI beyond the fundamental concerns of data
privacy and security. Five analytical aspects should be used to administer data governance in the realm of Personal Health Information (PHI): the field of data, those who are engaged, the significance or utilization of the PHI, the governance objective, and the governance platform.

In \cite{khatri2010designing}, the authors examined the fundamental model of contemporary data
governance and emphasised five crucial data decision fields: data postulates,  metadata, accuracy, access management, and information management cycle. Whereas a composite synthesis of research paper
\cite{abraham2019data} evaluates 145 academic and practitioner papers related to data governance,
encompassing the period from 2001 to 2019. The second one suggests a pyramidal governance system that uses governance mechanisms to balance data, realm, and organizational capacity. These mechanisms are
all structured by organizational ethical and technological “antecedents” pre-data ingestion and impacted by
risk control and performance-driven “consequences” post hoc. A comparative data governance activities examination found in academic and practitioner articles is
conducted in \cite{alhassan2018data}. The analysis explored a total of 120 information management elements that are classified as domains of governance, action, and decision areas. Authors
\cite{cheong2007need} constructed an organization governance model derived from a case study, the proposed framework includes three levels and their interconnections with one another. At the strategic
level of an organization, a data governance council’s responsibilities include approving guidelines, coordinating business and data projects, and assessing budget requests for data-related projects. In addition, significant roles are played by data custodians and data stewards on the tactical
level. The significant data stakeholders from various categories (user groups) operate at the lowest
level. This model aids in comprehending what organisational layer data governance duties should work
on; but, it does not provide a way to set up data governance.

Furthermore, the study \cite{were2017toward} determines the condition, factors influencing, and
potential obstacles to data governance in Kenyan health professional regulating entities. The
primary focus of this paper is to construct a framework which can be applicable to develop an official data
governance initiative at these healthcare governing bodies. This particular work determined the quality of data
maintenance, attaining customer satisfaction, safeguarding security and control of the data, and reaching a state of operational effectiveness as the driving force behind data governance in the governing
authorities. These bodies encounter challenges such as insufficient knowledge of data governance, lack
of management ownership and backing, and constrained distribution of funds and resources, each functioning as a
barrier to efficient data administration. The scholarly article \cite{li2019framework} suggested a
comprehensive layout for the governance of big data drawn from an analysis of ten representative
case studies of medical information exchange organizations in China. The framework was condensed
into the following three domains: drive domain, capability domain and support domain. It also
incorporates a total of 12 elements, such as massive data strategy formation, legal and regulatory
aspects, business actions practices, assistance, big medical data maintenance authority, collection of data, preservation, process and analysis, usage, resource utilization, quality management, and data protection safeguards that pertain to each respective domain.

However,  other  studies  \cite{al2018exploring,dasgupta2019conceptual,fleissner2014importance,juddoo2018data,kariotis2020emerging,paparova2023data,perez2021digital} are also considered which are not directly
investigate the healthcare data governance activities, frameworks or elements but highlight the
needs and importance of data governance frameworks or activities focusing on privacy and security
perspective, challenges and risks associated with data governance regarding big data, involvement of
IoT devices, data access controls or ownership, regulating multiple data actors, data quality
dimension in healthcare etc. These studies provide the basic elements or activities that should be
part of the healthcare data governance frameworks

\section{Healthcare Data Privacy \& Security Framework}
\label{sec:DGCF}
\begin{figure}
 \centering
 \includegraphics[scale=0.8]{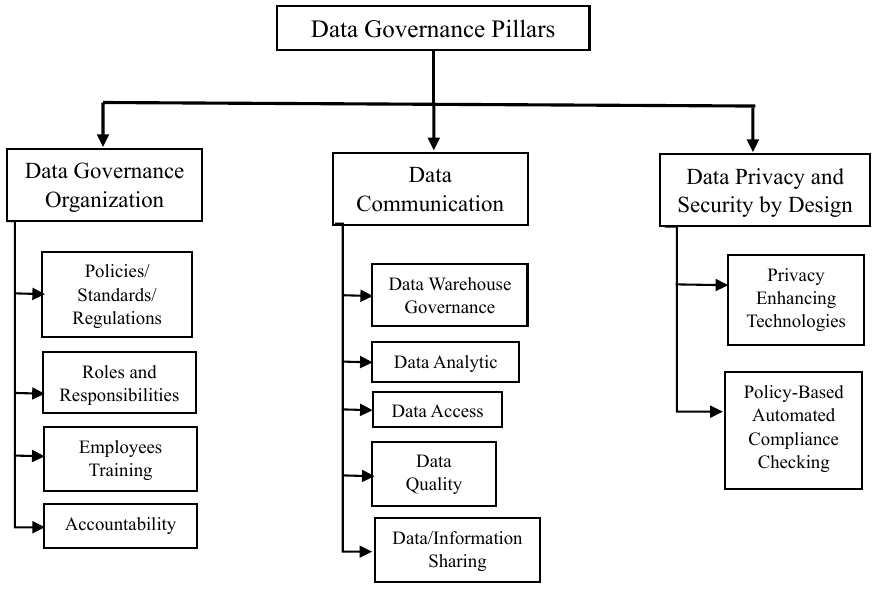}
 \caption {Healthcare Data Protection Governance - A Conceptual Framework}
 \label{fig:DGM}
\end{figure}

 These studies provide comprehensive data governance frameworks/elements/activities via which we
 categorised them into three distinct pillars: (1) Data Governance Organization, (2) Data
 Communication, and (3) Data Privacy and Security by Design, as shown in Figure \ref{fig:DGM}. This
 categorisation underscores healthcare data's security and privacy concerns throughout the chain,
 from records collection to sharing analysed results or data. However, existing state-of-the-art
 healthcare data governance frameworks treated privacy as an afterthought or an add-on feature, but
 instead as an integral part of the entire design process.
 
\subsection{Data Governance}
Data governance is critical to a successful data governance program. It provides a structured
approach to managing data by defining policies, rules, regulations, and procedures. The framework outlines the roles and responsibilities of various
stakeholders, including owners, stewards, custodians, and
users of data. Moreover, employees' training and accountability are also necessary and challenging
components of the data governance organization pillar in healthcare.

\subsubsection{Policies, Standards, Regulations:}
Fair    Information    Practice    Principles    (FIPPs)
\footnote{https://www.ftc.gov/sites/default/files/documents/reports/privacy-online-report-congress/priv-23a.pdf}
are a set of principles that form the basis of modern data protection laws and regulations. These
principles were first established by the United States in 1970s and have since been adopted by
many countries worldwide. The FIPPs provide a framework for gathering, utilizing, and revealing individual information fairly and transparently. In healthcare, the principles of FIPPs are reflected
in various regulations and acts, including the General Data Protection Regulation
(GDPR) \cite{lea2018will,regulation2018general} in the European Union, in the United States, the Health Insurance Portability and Accountability Act
(HIPAA) \cite{office2002standards}, the Personal Information
Protection and Electronic Documents Act (PIPEDA) \footnote{"Office of the Australian Information
 Commissioner (OIAC)", https://www.cdr.gov.au/} in Canada, etc. Additionally, healthcare
organizations must establish internal policies and regulations that provide a framework for data
privacy, security, consent management, and other critical aspects of healthcare compliance.

\subsubsection{Roles \& Responsibilities:}
Data governance requires collaboration and communication among various stakeholders to ensure data
is managed effectively, efficiently and securely. The roles and responsibilities in data governance
may differ according to the size of the organization, structure, and industry. Data Governance
organisational structure describes typical roles: 1) the composition, charter, and leadership of the Data Governance Committee; 2) the tasks of other relevant Data Governance committees 3) positions in huge
organisations; links to other groups and entities and 4) top-level funding. In addition, the structure demonstrates how the Data Governance
process gives supervision to owners of data, administrators, guardians, IT staff, compliance officers, and users of data, who are typically involved in the data warehouse management. 

\subsubsection{Employees Training:}
Healthcare workers are less tech-savvy and have less understanding of the safety of data, and technologies related to healthcare lag well behind those of the financial sector. Consequently, an understanding of data protection concerns requires education and training. Staff training works well for handling low-tech breaches of data. Keeping end-users informed about medical provider policies and fundamental security precautions is a part of preserving the privacy of data. Accidental interference can be prevented by teaching staff members about security risks including tapping on email links, pressing the computer desk login credentials, and visiting unapproved web pages.

\subsubsection{Accountability:}
Systems for evaluation and surveillance must be put in place by healthcare businesses to make certain that staff members are abiding by the regulations of the company. They create guidelines to prevent security lapses and specify precisely how infractions will be handled. Healthcare data is ideal for the identity thief. Being proactive and putting plans into action is essential. Reactive actions, like playing catch-up after a security breach happens, are not an effective approach.

\subsection{Data Communication}
Organizing or listing data elements with their description and other valuable information
(metadata) enables numerous insights into the core data or business concepts and
terminologies. Everyone implicated in the data management cycle uses the same terms to discuss the same
things, making communication easy. Effective communication reduces operational friction and
minimises data misuse due to inaccurate understanding. We further categorise this pillar into five
components to keenly analyse the security and privacy situation in the healthcare industry.

\subsubsection{Data Warehouse Governance (DWG):}
Data Warehouse Governance gives guidelines, rules and practices to guarantee gain,
utility, significance and risk management. Strategic choice-making and monitoring should ideally be the main concern of data warehouse governance, with secondary objectives including resource distribution, value of investment, and minimizing risk. Nevertheless, DWG could be more concerned regarding security and confidentiality, compliance, and risk prevention in healthcare settings given the delicate nature of protected health information.
\subsubsection{Data Analytic:}
Since the end of the 1950s, interest in artificial intelligence (AI) has cycled through periods of hope and disappointment due to unsatisfactory performance of algorithms and computing infrastructure
\cite{salathe2018focus}. While, the development of big data, machine learning, deep learning algorithms, and suitable computing infrastructure has rekindled interest in artificial intelligence (AI) technology and expedited its uptake across a number of industries. Even though modern AI techniques like machine learning have just recently been used in the healthcare industry, the prospects for better healthcare outcomes are encouraging.
\cite{whittlestone2019ethical}. Analyzing, or mining, the data without disclosing personally identifying or delicate private data about specific persons is known as privacy-preserving data analytics.

\subsubsection{Data Access:}
Institutions and people are both impacted by breaches of data and fraud. To safeguard patients' confidential information, the majority of healthcare organizations have developed sophisticated security protocols.
\cite{greenaway2015company}. While, software flaws, phishing, identities being stolen, and fraudulent attempts allow cyber criminals to get access and capture private information, which can lead to theft of identity, financial loss, anxiety, depression, prejudice, humiliation, assault, and other issues. \cite{chabot2015ontology,van2006challenge}. Insider dangers, however, are far more challenging to identify and stop than external ones. {\em Insider threat}
is any harm carried out by users with authorized access to an organization's network, apps, or data repositories. Therefore, strong security management structures are needed to guard against unauthorized access to healthcare information by outside parties and malevolent insider threats.

\subsubsection{Data Quality:}
Healthcare data quality is critical for delivering safe, efficient, and effective patient care,
driving medical research and innovation, informing healthcare policies, facilitating quality
improvement initiatives, ensuring interoperability, and complying with legal and regulatory
requirements. Governing bodies should design a data quality management framework composed of six key
dimensions: accuracy, completeness, consistency, uniqueness, timelessness and validity because
privacy and security compliance can only be achieved with accurate and consistent data.

\subsection{Data Privacy \& Security by Design}
The third pillar of the proposed data governance framework focuses on data fortification by design. Data Privacy and Security by Design (PSbD) is an approach to developing systems, products,
and services with a strong focus on privacy and security. Integrating privacy and security
considerations into any system's design, development, and implementation stages. After reviewing
the legal perspective through law articles, the privacy-by-design framework has been defined. By
doing so, medical data privacy risks can be minimized, and individuals can have more control over
their personal and sensitive data.

\subsubsection{Privacy Enhancing Technologies:}
Privacy-enhancing technologies (PETs) are tools, techniques, and systems designed to protect and
enhance privacy in the digital realm. They aim to safeguard sensitive information, limit data
collection and sharing, and give individuals greater control over their data. PETs can be applied in
various contexts of the data life-cycle and help healthcare providers meet their legal and ethical
obligations regarding data protection. 

\subsubsection{Policy-Based Automated Compliance Checking:}
Automated compliance checking involves leveraging technology, such as software applications or
platforms, to systematically evaluate the organization’s activities, processes, and behaviour to
comply with the established policies. To identify deviations or violations, these automated systems
can analyze various data sources, including logs, records, and transactions. As regulations and
policies evolve, the systems can be updated with new rules and requirements. This adaptability can
enable healthcare providers to stay compliant with the latest industry standards and regulations,
even in a rapidly changing healthcare landscape. 

\section{Conclusion}
The proliferation of data has brought about profound changes across various domains. Likewise, using cutting-edge technology for Electronic Medical Records not only transforms illness treatment but also benefits insurance, law enforcement, pharmaceutical, and other product-selling businesses. However, the healthcare industry is still grappling with persistent concerns related to data protection. It is noteworthy that mishandling healthcare data can lead to
significant liability, impacting patients and organisations. Healthcare entities must adhere to a
multitude of regulations to safeguard patient data. After thoroughly examining the literature, we
have determined that the previously established cutting-edge healthcare data governance frameworks
often regarded privacy as a secondary consideration rather than an inherent and essential component
integrated into the design process. We have conducted an in-depth review of existing studies and proposed a conceptual healthcare data governance framework with a primary focus on data privacy and security. 

\bibliographystyle{splncs04}
\bibliography{References}

\end{document}